\documentclass[%
 reprint,
superscriptaddress,
showpacs,
 amsmath,amssymb,
 aps,
prb,
floatfix,
]{revtex4-1}
\usepackage{graphicx,xcolor,wrapfig}
\usepackage{dcolumn}
\usepackage{bm}

\usepackage{ulem} 

\usepackage[colorlinks=true,citecolor=blue,linkcolor=blue]{hyperref}

\def\tred{\textcolor{red}}
\def\tblue{\textcolor{blue}}
\def\tcyan{\textcolor{cyan}}
\def\tgreen{\textcolor{green}}

\begin{document}


\title{Exciton-assisted low-energy magnetic excitations in a photoexcited Mott insulator on a square lattice}

\author{Kenji Tsutsui}
\email{tsutsui.kenji@qst.go.jp}
\affiliation{Synchrotron Radiation Research Center, National Institutes for Quantum Science and Technology, Hyogo 679-5148, Japan}

\author{Kazuya Shinjo}
\affiliation{Computational Quantum Matter Research Team, RIKEN Center for Emergent Matter Science (CEMS), Wako, Saitama 351-0198, Japan}

\author{Shigetoshi Sota}
\affiliation{Computational Materials Science Research Team, RIKEN Center for Computational Science (R-CCS), Kobe, Hyogo 650-0047, Japan}

\author{Takami Tohyama}
\email{tohyama@rs.tus.ac.jp}
\affiliation{Department of Applied Physics, Tokyo University of Science, Tokyo 125-8585, Japan}

\date{July 7, 2022; Accepted: February 24, 2023}
             

\begin{abstract}

The photoexcitation of a Mott insulator on a square lattice weakens the intensity of both single- and two-magnon excitations as observed in time-resolved resonant-inelastic X-ray scattering and time-resolved Raman scattering, respectively.  However, the spectral changes in the low-energy regions below the magnons have not yet been clearly understood.  To uncover the nature of the photoinduced low-energy magnetic excitations of the Mott insulator, we numerically investigate the transient magnetic dynamics in a photoexcited half-filled Hubbard model on a square lattice.  After turning off a pump pulse tuned for an absorption edge, new magnetic signals clearly emerge well below the magnon energy in both single- and two-magnon excitations.  We find that the low-energy excitations are predominantly created via excitonic states at the absorption edge. These exciton-assisted magnetic excitations may provide a possible explanation for the low-energy spectral weight in a recent time-resolved two-magnon Raman scattering experiment for insulating YBa$_2$Cu$_3$O$_{6.1}$. 
\end{abstract}
\maketitle


\section{Introduction}
\label{Sec1}

Photoirradiation of Mott insulators on a square lattice induces drastic changes in the electronic states, including a photoinduced insulator-to-metal transition~\cite{Okamoto2011,Hsieh2012,Li2022}.
The suppression of absorption spectral weights across the Mott-gap and the emergence of low-energy charge excitations inside the Mott-gap in optical conductivity are indications of a photoinduced insulator-to-metal transition.
In addition to the charge channel, the spin channel can also produce characteristic changes.
Spin dynamics in insulating antiferromagnet TbMnO$_3$~\cite{Bowlan2016} and KNiF$_3$~\cite{Batignani2015} have been observed in the optical pump, terahertz-probe spectroscopy, and femtosecond stimulated Raman scattering experiments, respectively.
In antiferromagnetic Mott insulators, a decrease of spectral weights in single-magnon dispersion owing to photoirradiation has been observed in time-resolved resonant-inelastic X-ray scattering (trRIXS) from Sr$_2$IrO$_4$~\cite{Dean2016} and Sr$_3$Ir$_2$O$_7$~\cite{Mazzone2020}, where RIXS can detect spin-flip excitations when incident X-rays are tuned to the $L$ edge in transition metals~\cite{Ament2011}.
A decrease in two-magnon weights has been observed in time-resolved two-magnon Raman scattering (trTMR) from an antiferromagnetic Mott insulator YBa$_2$Cu$_3$O$_{6.1}$~\cite{Yang2020}.
These decreases are naturally understood as a result of the emergence of photoexcited electronic states, which reduces antiferromagnetic spin correlation as proposed theoretically~\cite{Secchi2013,Mentink2017}. 

An emerging question is whether low-energy magnetic excitations below the magnon energy arise in the spin channel.
In fact, trTMR experiments have reported an increase of low-energy spectral weights below the two-magnon energy of $\sim$1000 to 1500~cm$^{-1}$ for photoirradiated YBa$_2$Cu$_3$O$_{6.1}$~\cite{Yang2020}.
A possible interpretation of the weight increase would be a shift of weight from the two-magnon peak due to the broadening of the peak.
Because photoirradiation to Mott insulators drastically changes the electronic states as seen in the charge channel, it is natural to expect that the spin channel will also induce low-energy excitations below magnon energies.

In this communication, we theoretically propose photoinduced low-energy magnetic excitations below the single- and two-magnon energies in a photoexcited Mott insulator on a square lattice.
Using a numerically exact-diagonalization (ED) technique and time-dependent density-matrix renormalization group (tDMRG) for a photoexcited half-filled Hubbard model on a square lattice, we find that low-energy magnetic excitations are induced by a pump pulse, whose intensity is maximized when the frequency of the pulse is tuned to the absorption edge.
By analyzing the low-energy magnetic excitation using point-group symmetry for the square lattice, we demonstrate that the photoinduced low-energy signals are predominantly created via photoexcited states with the $E$ representation of the $C_{4\mathrm{v}}$ point group, which are excitonic states at the absorption edge~\cite{Shinjo2021}.
The proposed exciton-assisted photoinduced magnetic excitations provide one of possible origins of low-energy weight in the trTMR spectrum.
This theoretical prediction will be confirmed as the pumping frequency is varied in the trTMR experiment.

\section{Results}
\label{Sec2}

\subsection{Model}

To describe Mott insulating states on a square lattice, we consider a single-band Hubbard model at half-filling, given by
\begin{equation}
H_0=-t_\mathrm{hop}\sum_{i\delta\sigma} c^\dagger_{i\sigma} c_{i+\delta\sigma}
 + U\sum_i n_{i\uparrow}n_{i\downarrow},
\label{singleH}
\end{equation}
where $c^\dagger_{i\sigma}$ is the creation operator of an electron with spin $\sigma$ at site $i$ and number operator $n_{i\sigma}=c^\dagger_{i\sigma}c_{i\sigma}$, $i+\delta$ represents the four nearest-neighbor sites around site $i$.
$t_\mathrm{hop}$ and $U$ are the nearest-neighbor hopping and on-site Coulomb interactions, respectively.
We take $U/t_\mathrm{hop}=10$, which guarantees antiferromagnetic Mott insulating ground state at half filling.
Note that $t_\mathrm{hop}\sim 0.35$~eV for cuprates. 

We incorporate an external electric field via the Peierls substitution in the hopping term, $c^\dagger_{i,\sigma} c_{i+\delta,\sigma} \rightarrow e^{i\mathbf{A}(t)\cdot\mathbf{R}_\delta} c^\dagger_{i,\sigma} c_{i+\delta,\sigma}$, leading to a time-dependent driven Hamiltonian $H(t)$.
Here, $\mathbf{R}_\delta$ is the vector from $i$ to $i+\delta$, and $\mathbf{A}(t)$ at time $t$ is the vector potential, given by
$\mathbf{A}(t)=\mathbf{A}_0 e^{-(t-t_0)^2/(2t_d^2)} \cos[\omega_\mathrm{p}(t-t_0)]$, 
where a Gaussian-like envelope centered at $t=t_0$ has temporal width $t_d$ and a central frequency $\omega_\mathrm{p}$.
We apply an external electric field along the $x$ direction without otherwise specifications, that is, $\mathbf{A}_0=(A_0,0)$, and set $A_0=0.5$, $t_0=0$, and $t_d=0.5$.
Hereafter we use $t_\mathrm{hop}=1$ as the energy unit and $1/t_\mathrm{hop}$ as the time unit. 

For calculating the time-resolved spin excitation in trRIXS during pumping, a real-time representation of the cross-section was used in refs.~\cite{Wang2018,Wang2021} for a Hubbard model on a square lattice.
In contrast, we focus on time-resolved spin excitations after pumping.
In this case, it is convenient to use the time-dependent wave function $\left|\psi\left(t\right)\right>$ as the initial state of the dynamical spin susceptibility for a time period after turning off the pump pulse, $t>t_\mathrm{off}$, when the Hamiltonian is time-independent.
Applying this procedure, we obtain the time-resolved dynamical susceptibility with momentum $\mathbf{q}$ and frequency $\omega$ for a $\mathbf{q}$-dependent physical quantity $O_\mathbf{q}$ as~\cite{Shinjo2017}
\begin{eqnarray}
&O&\left({\bf q},\omega;t\right)\nonumber\\
&=&\frac{1}{\pi}\mathrm{Re}\int_t^\infty ds e^{i\left(\omega+i\eta\right)s} \left<\psi\left(t\right)\right| \left[O_{\bf q}(s),O_{\bf -q}\right] \left|\psi\left(t\right)\right> \label{O1}\\
&=&-\frac{1}{\pi}\mathrm{Im}\sum_{m,n}\frac{1}{\omega-(\varepsilon_n-\varepsilon_m)+i\eta}
\nonumber\\
&&\times\left[
\left<\psi\left(t\right)|m\right>
\left<m\right|O_{\bf q}\left|n\right>
\left<n\right|O_{\bf -q}\left|\psi\left(t\right)\right>
\right.\nonumber\\
&&-\left.
\left<\psi\left(t\right)\right|O_{\bf -q}\left|m\right>
\left<m\right|O_{\bf q}\left|n\right>
\left<n|\psi\left(t\right)\right>
\right],
\label{O2}
\end{eqnarray}
where the operator $O_{\bf q}(s)=e^{iH_0s}O_{\bf q}e^{-iH_0s}$, $H_0\left| m\right>=\varepsilon_m\left| m\right>$, $\eta$ is a small positive number, and $\left|\psi\left(t\right)\right>=e^{-iH_0(t-t_\mathrm{off})}\left|\psi\left(t_\mathrm{off}\right)\right>$.
Note that replacing $\left|\psi\left(t\right)\right>$ with the ground state $\left|0\right>=\left|\psi\left(-\infty\right)\right>$ in Eq.~(\ref{O1}) formally gives the equilibrium dynamical susceptibility $O(\mathbf{q},\omega)$.
For an $L$-site periodic lattice, we choose $O_\mathbf{q}=S^z_\mathbf{q}=L^{-1/2}\sum_i e^{-i\mathbf{q}\cdot\mathbf{R}_i} S^z_i$ for the time-resolved dynamical spin susceptibility $S(\mathbf{q},\omega;t)$ and $O_\mathbf{q}=\sum_\mathbf{k}\left(\cos k_x-\cos k_y\right) \mathbf{S}_{\mathbf{k}+\mathbf{q}}\cdot\mathbf{S}_{-\mathbf{k}}$ with $\mathbf{q}=0$ for the trTMR susceptibility $M(\omega;t)$ with $B_\mathrm{1g}$ representation, where $S^z_i$ is the $z$ component of spin operator $\mathbf{S}_i$ at site $i$.
The first term in Eq.~(\ref{O2}), denoted hereafter as $O_1\left({\bf q},\omega;t\right)$, corresponds to the time-resolved dynamical correlation function and is related to the second term, $O_2\left({\bf q},\omega;t\right)$, as $O_2\left({\bf q},\omega;t\right)=-O_1\left({\bf q},-\omega;t\right)$, where $O\left({\bf q},\omega;t\right)=O_1\left({\bf q},\omega;t\right)+O_2\left({\bf q},\omega;t\right)=O_1\left({\bf q},\omega;t\right)-O_1\left({\bf q},-\omega;t\right)$.
The integration of $S_1\left({\bf q},\omega;t\right)$ with respect to $\omega$ ($-\infty\leq\omega\leq\infty$) yields the time-resolved static spin structure factor $S(\mathbf{q};t)\equiv\left<\psi\left(t\right)\right|S^z_\mathbf{q} S^z_{-\mathbf{q}}\left|\psi\left(t\right)\right>$.
The time-dependent wave function $\left|\psi\left(t\right)\right>$ is determined by the procedure described in the Methods section.

\begin{figure*}[t]
\includegraphics[width=0.6\textwidth]{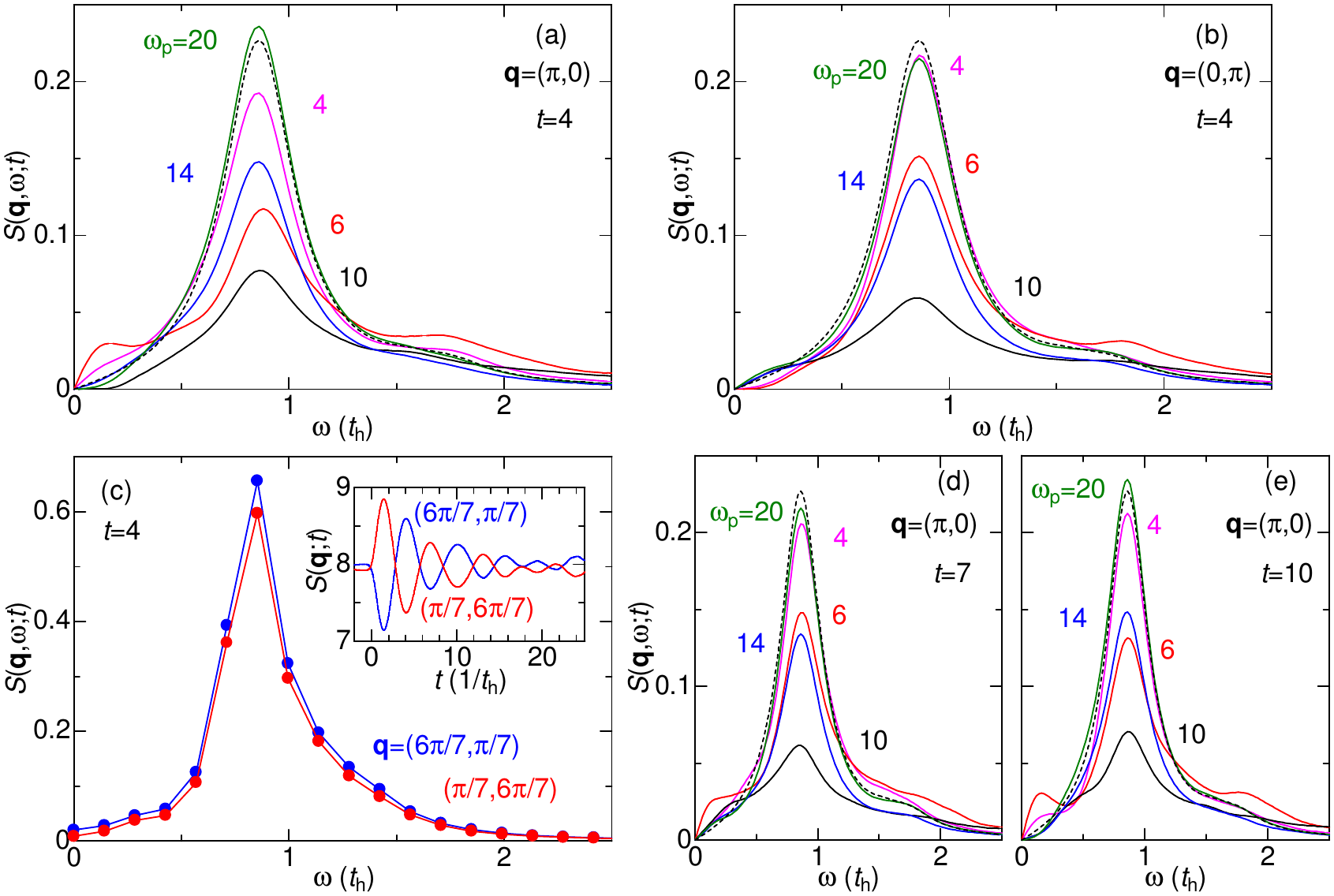}
\caption{{\bf Pumping frequency dependence of time-resolved dynamical spin susceptibility after pumping.}
A half-filled Hubbard model on square lattices with on-site Coulomb interaction $U=10$ is considered.
\textbf{a} Momentum $\mathbf{q}=(\pi,0)$ and \textbf{b} $\mathbf{q}=(0,\pi)$ for a periodic 4$\times$4 lattice obtained by the exact-diagonalization (ED) at $t=4$.
\textbf{c} Pumping frequency $\omega_\mathrm{p}=20$ and time $t=4$ for $\mathbf{q}=(\pi/7,6\pi/7)$ (red circles) and $\mathbf{q}=(6\pi/7,\pi/7)$ (blue circles) in a 6$\times$6 lattice with open boundary conditions obtained by the time-dependent density-matrix renormalization group.
The inset shows the time-resolved static spin structure factor $S(\mathbf{q};t)$.
The antiphase oscillation agrees with a previous result obtained by ED \cite{Tsutsui2021}.
\textbf{d} $t=7$ and \textbf{e} $t=10$ at $\mathbf{q}=(\pi,0)$ for the 4$\times$4 lattice calculated using ED.
In (\textbf{a, b, d, e}), dashed black lines represent the results before pumping, whereas the purple, read, black, blue, and green solid lines represent the results at $\omega_\mathrm{p}$=4, 6, 10, 14, and 20, respectively.}
\label{fig1}
\end{figure*}

\subsection{Low-energy excitation below a single magnon.}
We first present the results of $S(\mathbf{q},\omega;t)$, which can describe photoinduced low-energy magnetic excitations below the single-magnon dispersion energy.
For a 4$\times$4 lattice, we take $\mathbf{q}=(\pi,0)$ and $(0,\pi)$, where the energy of single magnon is maximized.
Figures~\ref{fig1}a, b show the pumping frequency $\omega_\mathrm{p}$ dependence of $S(\mathbf{q},\omega;t)$ at $\mathbf{q}=(\pi,0)$ and $(0,\pi)$, respectively, for $t=4$.
The peak at $\omega=0.8$ represents a single-magnon excitation at a given $\mathbf{q}$.
At a high pumping frequency $\omega_\mathrm{p}=20>U$, the peak intensity is larger and smaller than that before pumping (dotted lines) in Figs.~\ref{fig1}a, b, respectively.
This $\mathbf{q}$-dependent intensity at a given $t$ is a consequence of an antiphase oscillation of time-dependent spin structure factors in a photoexcited Mott insulator on a square lattice proposed by the present authors~\cite{Tsutsui2021} [see also the inset of Fig.~\ref{fig1}c].
With decreasing $\omega_\mathrm{p}$, the peak intensity decreases and reaches  a minimum at $\omega_\mathrm{p}=10$, where the energy absorbed by pumping in the system is the largest, leading to a weakening of antiferromagnetic spin correlation~\cite{Wang2018,Wang2021}.
With a further decrease in $\omega_\mathrm{p}$, the peak intensity increases.

At $\omega_\mathrm{p}=6$, which corresponds to the absorption peak energy at the Mott-gap edge~\cite{Tohyama2005}, a hump structure is observed at $\omega=0.2$ below the peak energy for $\mathbf{q}=(\pi,0)$.
This can be attributed to a photoinduced low-energy magnetic excitation.
An interesting observation is that, away from $\omega_\mathrm{p}=6$, the hump structure loses its weight.
In particular, when $\omega_\mathrm{p}$ satisfies the off-resonance condition, $\omega_\mathrm{p}=20$, the hump structure almost disappears.
This behavior may not be dependent on system size.
To confirm this, we perform tDMRG calculations of $S(\mathbf{q},\omega;t)$ for a 6$\times$6 lattice with open boundary conditions (see the Methods section), and $S(\mathbf{q},\omega;t=4)$ at $\mathbf{q}=(6\pi/7,\pi/7)$ and $\mathbf{q}=(\pi/7,6\pi/7)$ for $\omega_\mathrm{p}=20$ are shown in Fig.~\ref{fig1}c.
The time-resolved static spin structure factor $S(\mathbf{q};t=4)$ shows a larger value at $\mathbf{q}=(\pi/7,6\pi/7)$ than at $\mathbf{q}=(6\pi/7,\pi/7)$, as shown in the inset of Fig.~\ref{fig1}c.
The magnon peak at $\omega=0.6$ is more intense in intensity at $\mathbf{q}=(\pi/7,6\pi/7)$ than at $\mathbf{q}=(6\pi/7,\pi/7)$.
However, we find no clear hump at the low-energy region in either $\mathbf{q}$s, which is the same as the results at $\omega_\mathrm{p}=20$ for the 4$\times$4 periodic lattice.
This indicates that the effects of system size and boundary conditions are small. 

Interestingly, there is no hump structure at $\mathbf{q}=(0,\pi)$ in Fig.~\ref{fig1}b, for $\omega_\mathrm{p}=6$.
Because the pump pulse is polarized along the $x$ direction, it is meaningful to determine the difference of low-energy magnetic excitations between $\mathbf{q}=(\pi,0)$ and $(0,\pi)$.
This is discussed further in the Discussions section.

The $\omega_\mathrm{p}$ dependence of the low-energy hump structure for $\mathbf{q}=(\pi,0)$ remains unchanged at $t=7$ and $t=10$, as shown in Figs.~\ref{fig1}d, e, respectively.
This implies that the hump intensity does not oscillate with time, in contrast to the peak intensity~\cite{Tsutsui2021}.
Therefore, it is possible to detect this hump in the trRIXS of insulating cuprates and iridates on square lattices when a pump pulse tuned to a Mott-gap edge is applied along the $x$ direction and momentum transfer is set to $\mathbf{q}=(\pi,0)$.

\begin{figure}[t]
\includegraphics[width=0.45\textwidth]{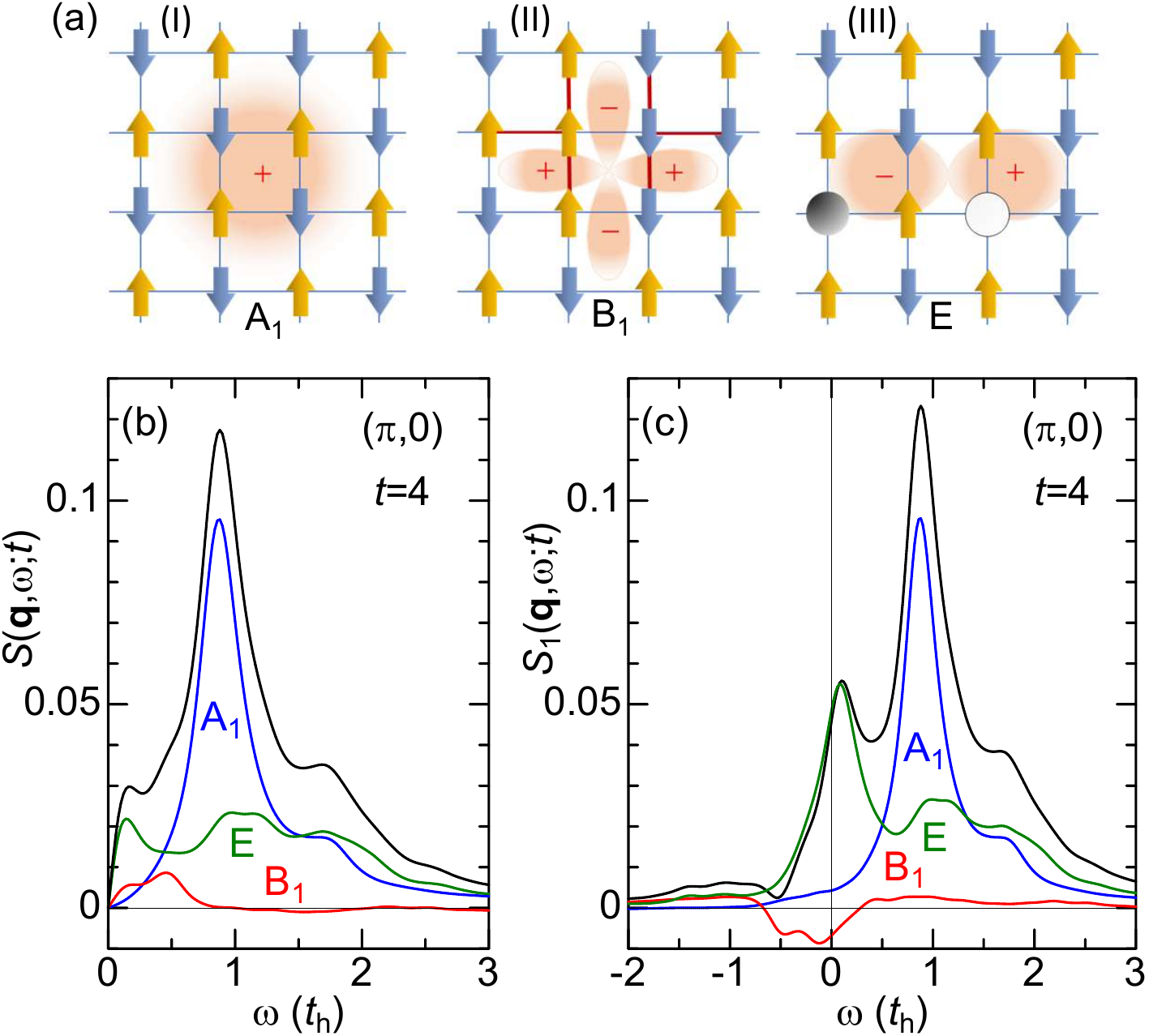}
\caption{{\bf Symmetry decomposition of the time-resolved wave function and the time-resolved dynamical spin susceptibility.}
Momentum $\mathbf{q}=(\pi,0)$ and the time $t=4$ at pumping frequency $\omega_\mathrm{p}=6$ for a half-filled 4$\times$4 periodic Hubbard lattice with on-site Coulomb interaction $U=10$.
\textbf{a} Schematic view of the wave function in terms of (I) $A_1$, (II) $B_1$, and (III) $E$ representations of the $C_\mathrm{4v}$ point group.
The up and down arrows denote up and down spins, respectively, located on the lattice sites.
The white (black) circle in the right panel represents a holon (doublon).
Because the $A_1$, $B_1$, and $E$ representations correspond to $s$, $d_{x^2-y^2}$, and $p_x$ waves, respectively, those waves are visualized as a wave-function form with signs in each panel.
\textbf{b} Time-resolved dynamical spin susceptibility $S(\mathbf{q},\omega;t)$ and \textbf{c} time-resolved dynamical correlation function $S_1(\mathbf{q},\omega;t)$.
Note that $S(\mathbf{q},\omega;t)=S_1(\mathbf{q},\omega;t)-S_1(\mathbf{q},-\omega;t)$.
The black solid lines depict the sum of each contribution (the blue, green, and red solid lines depict the contributions of $A_1$, $E$, and $B_1$ of the $C_\mathrm{4v}$ point group, respectively). }
\label{fig2}
\end{figure}

To understand the origin of the hump structure at $\omega_\mathrm{p}=6$, we decompose $S(\mathbf{q},\omega;t)$ at $\mathbf{q}=(\pi,0)$ and $t=4$ into several contributions.
This is logical step because $\left|\psi\left(t\right)\right>$ for $t>t_\mathrm{off}$ has three components arising from irreducible representations, $A_1$, $B_1$, and $E$, of the square lattice with the $C_\mathrm{4v}$ point group:
\begin{equation}
\left|\psi\left(t\right)\right>=\left|\psi_{A_1}\left(t\right)\right>+\left|\psi_{B_1}\left(t\right)\right>+\left|\psi_E\left(t\right)\right>.
\label{sym}
\end{equation}
Note that the $A_1$ and $B_1$ representations correspond to $s$ and $d_{x^2-y^2}$ waves, respectively, as schematically presented in Fig.~\ref{fig2}a.
The dominant contribution to the $A_1$ state in Eq.~(\ref{sym}) is numerically identified as the ground state, where spins predominately arrange antiferromagnetically.
The left panel in Fig.~\ref{fig2}a schematically represents this spin arrangement.
On the other hand, the dominant contribution to the $B_1$ state originates from the two-photon absorbed states, including the $B_{1g}$ Raman active state, where neighboring two spins are flipped, and two-magnons are excited as schematically shown in the middle panel of Fig.~\ref{fig2}a.
The $E$ representation corresponds to a $p_x$ wave, because the $E$ state agrees with optically allowed single-photon absorbed states created by an electric field along the $x$ direction.
In this $E$ state, a holon-doublon pair shown in the right panel of Fig.~\ref{fig2}a is created and the pair forms an excitonic state in order to gain the magnetic energy in the spin background~\cite{Shinjo2021}.
The spin arrangement of the $E$ state gives rise to the low-energy magnetic excitations, as discussed in the Discussions section.

Figure~\ref{fig2}b shows the three contributions to $S(\mathbf{q}=(\pi,0),\omega;t=4)$, where the $A_1$, $B_1$, and $E$ states in Eq.~(\ref{sym}) couple to the final states with $A_1$, $A_1$, and $B_1$, respectively, at $\mathbf{q}=(\pi,0)$.
We find that the magnon peak consists mainly of the $A_1$ state in Eq.~(\ref{sym}).
This is reasonable because of the presence of a significant ground-state contribution in the $A_1$ state.
However, we find that the E state has the largest contribution at approximately $\omega=0.2$. 

$S(\mathbf{q},\omega;t)$ has two contributions from two terms in Eq.~(\ref{O2}).
Let the first term be denoted by $S_1(\mathbf{q},\omega;t)$ and the second term by $S_2(\mathbf{q},\omega;t)$.
It is crucial to identify which term dominates low-energy excitations.
Figure~\ref{fig2}c shows $S_1(\mathbf{q},\omega;t)$ at $\mathbf{q}=(\pi,0)$ decomposed into the three irreducible representations.
For $\omega\ge0$, $S(\mathbf{q},\omega;t)=S_1(\mathbf{q},\omega;t)-S_1(\mathbf{q},-\omega;t)$.
This explains why the spectral weight at $\omega=0$ is exactly zero in $S(\mathbf{q},\omega;t)$.
We find that the $E$ component in $S_1(\mathbf{q},\omega;t)$ has a large weight at $\omega=0.1$.
This weight results in the $\omega=0.2$ hump structure in $S(\mathbf{q},\omega;t)$.
Such a large weight near $\omega=0$ in $S_1(\mathbf{q},\omega;t)$ indicates the presence of very low-energy magnetic excitations from the single-photon absorbed $E$ state at the Mott-gap edge, as will be discussed in the Discussions section.

\begin{figure}[t]
\includegraphics[width=0.4\textwidth]{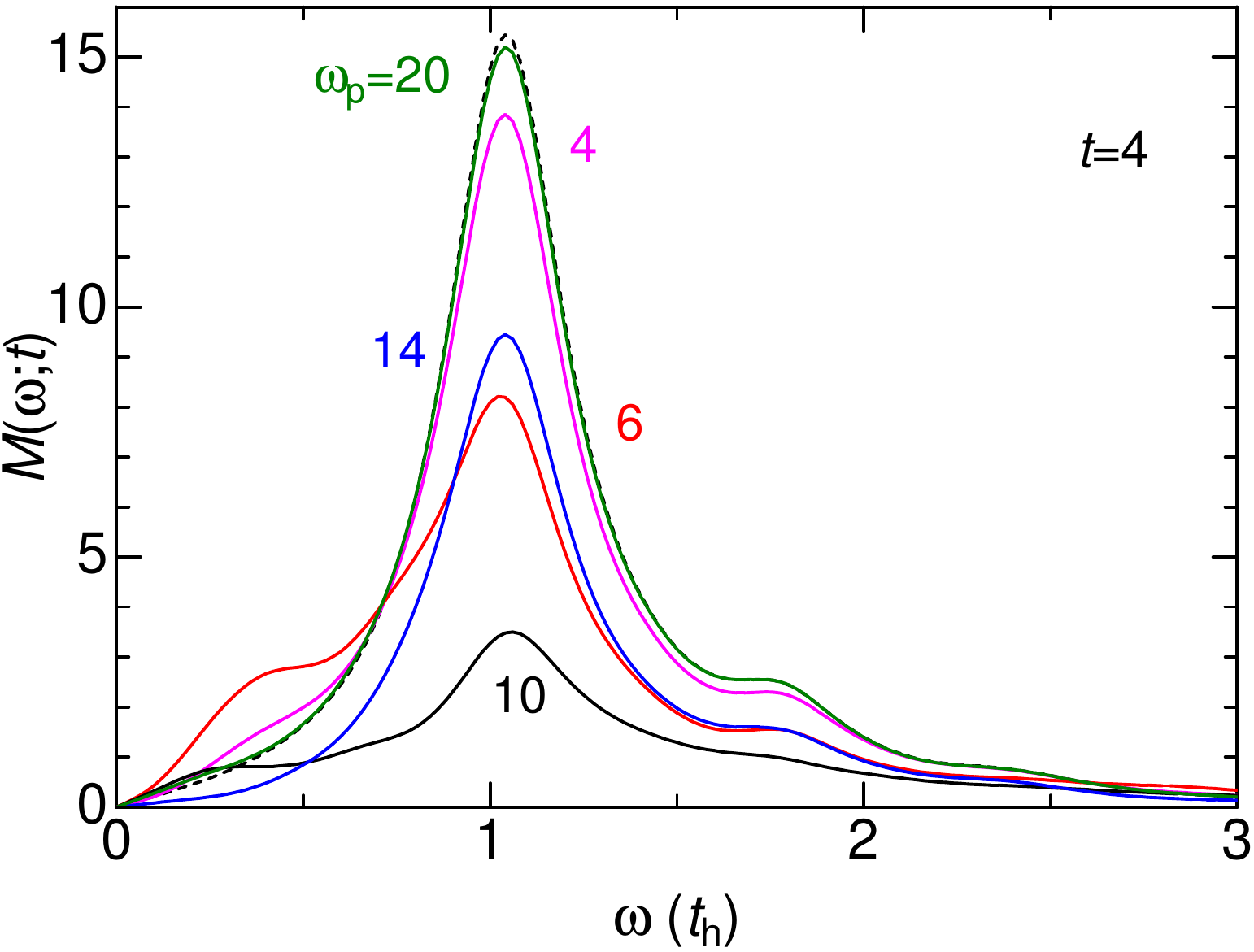}
\caption{{\bf Pumping frequency dependence of time-resolved two-magnon Raman susceptibility  after pumping.}
Time $t=4$ for a half-filled 4$\times$4 Hubbard lattice with on-site Coulomb interaction $U=10$.
The dashed black line represents the results before pumping, while purple, read, black, blue, and green solid lines represent the results at pumping frequency $\omega_\mathrm{p}$=4, 6, 10, 14, and 20, respectively.}
\label{fig3}
\end{figure}

\subsection{Low-energy excitation below two magnons.}
Photoirradiation may induce low-energy magnetic excitations below the two-magnon energy.
To confirm this, we perform ED calculations of the trTMR susceptibility after turning off the pump pulse using a 4$\times$4 periodic half-filled Hubbard lattice.
Figure~\ref{fig3} shows the $\omega_\mathrm{p}$ dependence of $M(\omega;t)$ at $t=4$.
The pump pulse suppresses the intensity of a two-magnon peak at $\omega=1$, which has been observed in the trTMR experiment for photoirradiated YBa$_2$Cu$_3$O$_{6.1}$~\cite{Yang2020}.
The peak-intensity variation with respect to $\omega_\mathrm{p}$ is similar to single-magnon intensity, as shown in Figs.~\ref{fig2}b, c.
The peak position for each $\omega_\mathrm{p}$ is almost the same as the position before pumping, and the line shape of the peak near the maximum height can be fitted by a single Lorentzian as in the equilibrium case.
These facts are consistent with the significant contribution of the ground state in $A_1$ component of the time-dependent wave function when $A_0=0.5$, as discussed in the "Low-energy excitation below single magnon" subsection in the Results.
With increasing $A_0$, the $A_1$ contribution decreases and the $B_1$ and $E$ contributions increase, which leads to an anisotropic line shape (see Supplementary Note).
In the experiments, even for an equilibrium case before pumping, an asymmetric line shape in the two-magnon peak was reported and its origin was attributed to electron-phonon interactions~\cite{Farina2018}, which are not included in our calculations.

\begin{figure}[t]
\includegraphics[width=0.4\textwidth]{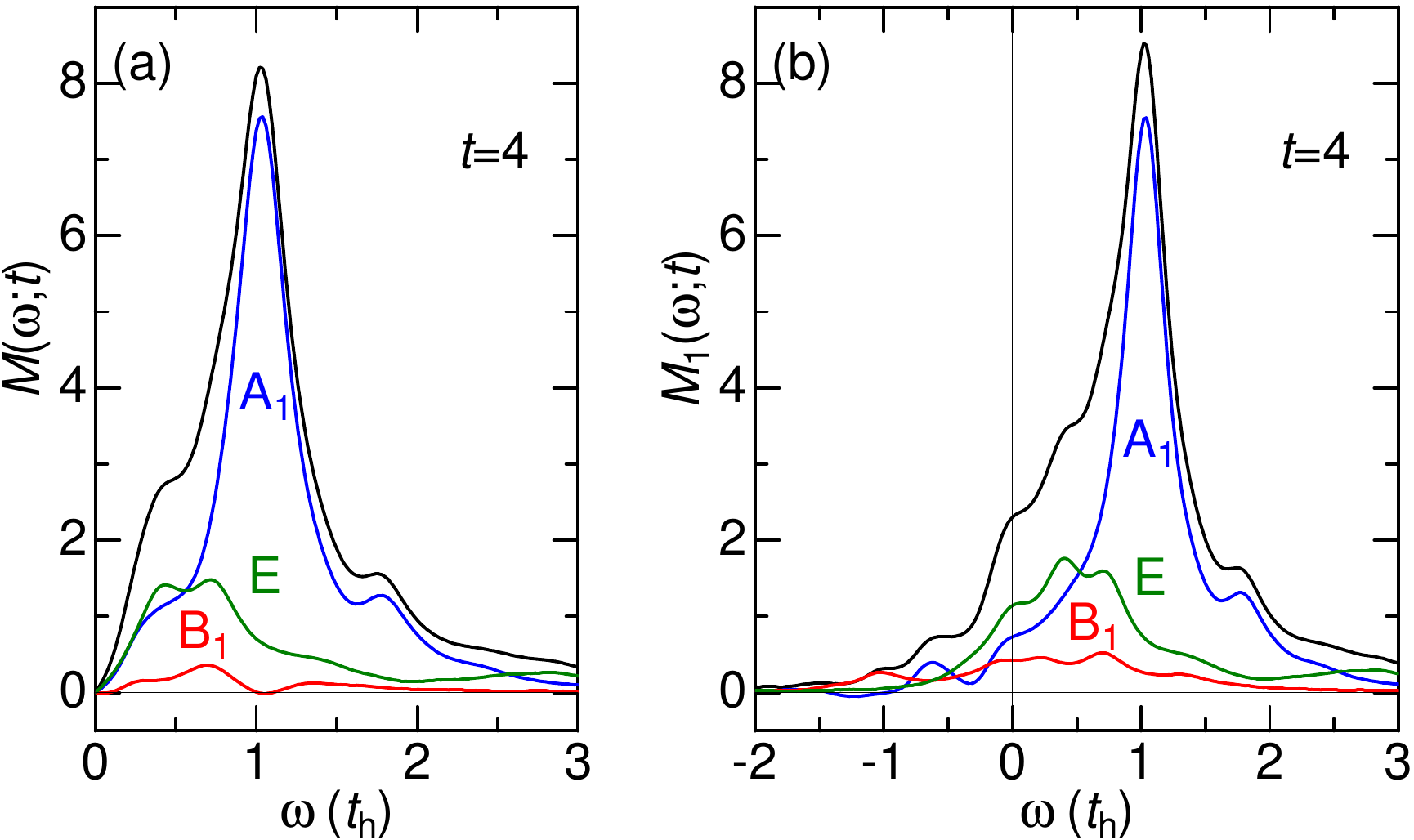}
\caption{{\bf Symmetry decomposition of time-resolved two-magnon Raman spectra.}
Time $t=4$ at pumping frequency $\omega_\mathrm{p}=6$ for a half-filled 4$\times$4 periodic Hubbard lattice with on-site Coulomb interaction $U=10$.
\textbf{a} Time-resolved two-magnon Raman susceptibility $M(\omega;t)$ and \textbf{b} time-resolved dynamical two-magnon Raman correlation function $M_1(\omega;t)$.
Note that $M(\omega;t)=M_1(\omega;t)-M_1(-\omega;t)$.
The black solid lines depict the sum of each contribution (the blue, green, and red solid lines depict the contributions of $A_1$, $E$, and $B_1$ of the $C_\mathrm{4v}$ point group, respectively).}
\label{fig4}
\end{figure}

When $\omega_\mathrm{p}=6$, we observe an enhancement in the low-energy spectral weight at approximately $\omega=0.4$.
This is similar to the enhancement of low-energy weight in $S(\mathbf{q},\omega;t)$ shown in Fig.~\ref{fig1}a, although its energy in $S(\mathbf{q},\omega;t)$ is nearly a half of $\omega=0.4$.
This similarity suggests same origin for these enhancements.
To identify the symmetry components that contribute to the enhancement in trTMR, we decompose the spectral weight into several components characterized by irreducible representations, as in Fig.~\ref{fig2}.
These are shown in Fig.~\ref{fig4}.
Similar to $S(\mathbf{q},\omega;t)$, the low-energy part of $M(\omega;t)$ originates from the $E$ component, as shown in Fig.~\ref{fig4}a.
From Fig.~\ref{fig4}b, the $E$ component corresponds to the first term in Eq.(\ref{O2}), i.e.,  $M_1(\omega;t)$.
In other words, the low-energy excitations at $\omega=0.4$ in $M(\omega;t)$ is a consequence of magnetic excitations from a single-photon absorbed state at the Mot-gap edge. 

\begin{figure}[t]
\includegraphics[width=0.4\textwidth]{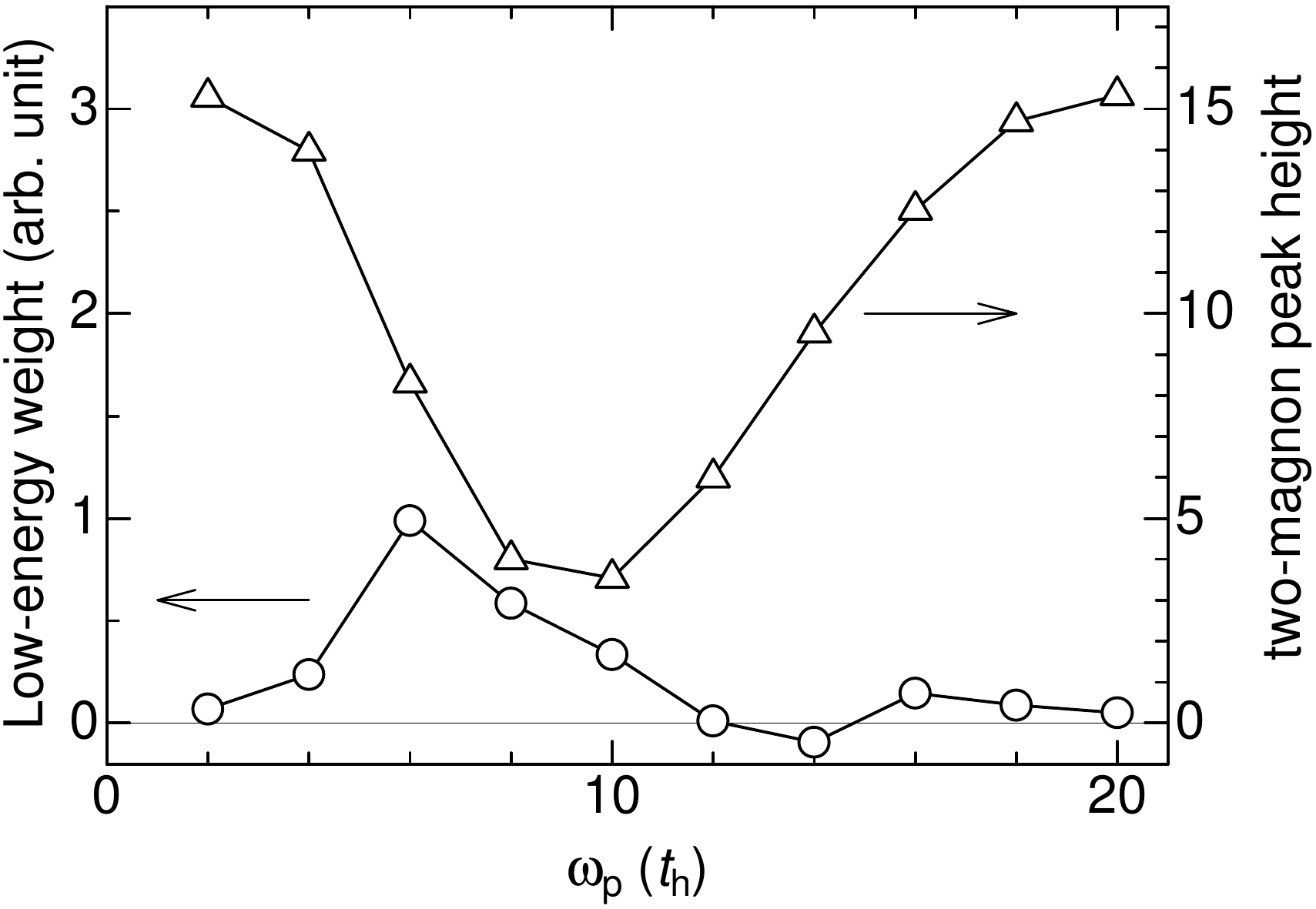}
\caption{{\bf Pumping frequency dependence of time-resolved two-magnon Raman susceptibility.}
Low-energy spectral weight below frequency $\omega=0.7$ (open circles) and height of two-magnon peak (open triangles) obtained from time-resolved two-magnon Raman susceptibility $M(\omega; t=4)$ shown in Fig.~\ref{fig3}.}
\label{fig5}
\end{figure}

To investigate the $\omega_\mathrm{p}$ dependence of trTMR in more detail, we analyze low-energy spectral weight below $\omega=0.7$ in $M(\omega;t=4)$ shown in Fig.~\ref{fig3}.
The weight is evaluated by subtracting the contribution of the two-magnon peak fitted by a single Lorentzian.
Figure~\ref{fig5} shows the low-energy weight and the two-magnon peak height  as functions of $\omega_\mathrm{p}$.
The peak height decreases with increasing $\omega_\mathrm{p}$ and exhibits a minimum at $\omega_\mathrm{p}=10$ corresponding to the center of the optical absorption spectrum for $U=10$~\cite{Tohyama2005}.
In contrast, the low-energy weight exhibits a maximum at $\omega_\mathrm{p}=6$ corresponding to the Mott-gap edge.
The difference in $\omega_\mathrm{p}$, which shows the peak height minimum and low-energy weight maximum, clearly demonstrates that the enhancement of low-energy excitation is not due to spectral weight transfer from the two-magnon peak.
Based on this result, we can predict different $\omega_\mathrm{p}$ dependences between the low-energy weight and peak height in the trTMR for Mott insulators such as YBa$_2$Cu$_3$O$_{6.1}$.

\section{Discussions}
\label{Sec3}
In the Results section, we have observed that low-energy magnetic excitations are induced in both $S(\mathbf{q},\omega;t)$ and $M(\omega;t)$ when the pumping frequency $\omega_\mathrm{p}$ is tuned to the Mott-gap edge at $\omega=6$~\cite{Tohyama2005}.
We have also found that the state with the $E$ representation in $\left|\psi\left(t\right)\right>$ contributes most significantly to these excitations, as shown in Figs.~\ref{fig2}, \ref{fig4}.
To understand these facts, we should notice that there is an excitonic peak at the Mott-gap edge in the optical conductivity~\cite{Shinjo2021,Tohyama2005}.
The excitonic peak is the optically allowed state with the $E$ representation and originates not from long-range Coulomb interactions but from magnetic effects gaining the magnetic energy in the spin background.
Therefore, it is reasonable to assume that the low-energy magnetic excitations are related to the excitonic state.
We separately confirmed that the $E$ component of $S_1(\mathbf{q}=(\pi,0),\omega;t=4)$ in Fig.~\ref{fig2}c [$M_1(\omega;t=4)$ in Fig.~\ref{fig3}b] is almost equivalent to the equilibrium $S(\mathbf{q}=(\pi,0),\omega)$ [$M(\omega)$] obtained by considering the excitonic state as the initial state.
This is strong evidence that the photoinduced magnetic excitations are assisted by the excitonic state.  

\begin{figure}[t]
\includegraphics[width=0.4\textwidth]{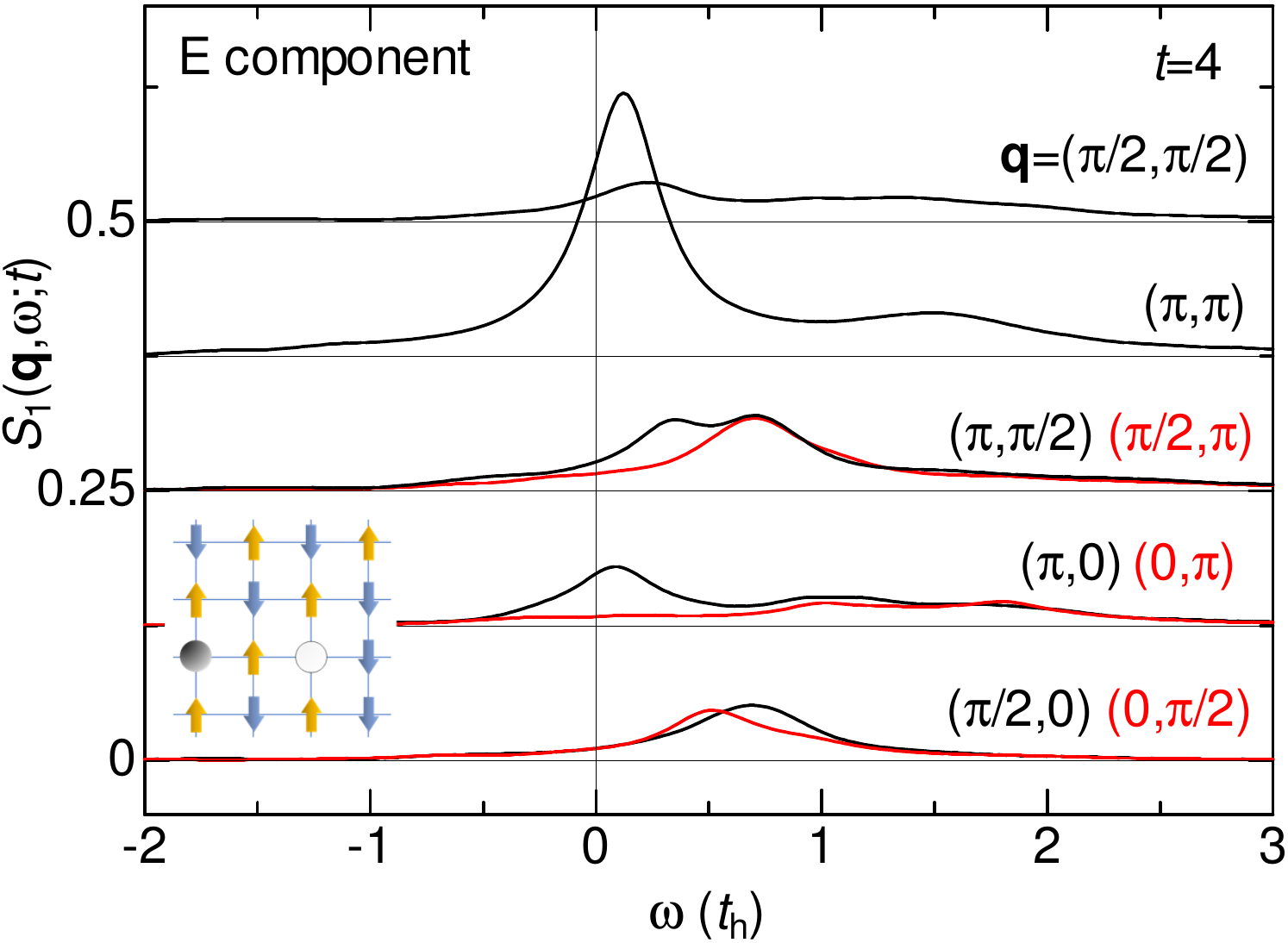}
\caption{{\bf Momentum dependence of symmetry-decomposed E component of time-resolved dynamical spin correlation function.}
Time $t=4$ at pumping frequency $\omega_\mathrm{p}=6$ for a half-filled 4$\times$4 periodic Hubbard lattice with on-site Coulomb interaction $U=10$.
The results at momentum $\mathbf{q}=(\pi,0)$ is the same as those in Fig.~\ref{fig2}c.
The inset shows a configuration with the largest weight in an $E$-symmetry excitonic state at the Mott-gap edge that contributes to $\left|\psi_\mathrm{E}\left(t\right)\right>$ in Eq.~(\ref{sym}).
The white (black) circle represents a holon (doublon) surrounded by up and down spins that arrange antiferromagnetically along the $x$ direction but exhibit lower antiferromagnetic correlation along the $y$ direction.}
\label{fig6}
\end{figure}

The $E$ component below the main peak of $M_1(\omega;t=4)$ in Fig.~\ref{fig4}b is wider than that of $S_1(\mathbf{q}=(\pi,0),\omega;t=4)$ in Fig.~\ref{fig2}c.
A possible explanation of the wide distribution in $M_1(\omega;t)$ is that two-magnon excitations are a combination of a pair of single magnons with opposite momenta; that is, $M_1(\omega;t)$ composed of the sum of $S_1(\mathbf{q},\omega;t)$ with different momenta.
To elucidate this wide distribution of the $E$ component in $M_1(\omega;t=4)$ for $\omega_\mathrm{p}=6$, we calculate the $E$ component of $S_1(\mathbf{q},\omega;t=4)$ for all $\mathbf{q}$.
The $\mathbf{q}$ dependence is shown in Fig.~\ref{fig6}.
The result for $\mathbf{q}=(\pi,0)$ is shown in Fig.~\ref{fig2}c.
We find that the $\mathbf{q}=(\pi,0)$ and $(\pi,\pi)$ excitations have the lowest energy, and other $\mathbf{q}$s have higher excitation energies.
Because many pairs of single magnons with higher energies, such as a pair at $\mathbf{q}=(\pm\pi/2,0)$, contribute to $M_1(\omega;t)$, the distribution of the $E$ component in $M_1(\omega;t=4)$ becomes wider.
Furthermore, Fig.~\ref{fig6} shows that the photoinduced magnetic excitations in $S(\mathbf{q}=(\pi,0),\omega;t=4)$ are lower in energy than those of $M(\omega;t=4)$. 

Finally, we comment on why the $\mathbf{q}=(\pi,0)$ excitation is the lowest energy in Fig.~\ref{fig6}.
This may be due to the magnetic configurations in the excitonic state.
In this state, a holon-doublon pair is surrounded by localized spins in the background.
The pair is created along the $x$ direction parallel to an applied electric field, as shown in the inset of Fig.~\ref{fig6} (see also the right panel of Fig.~\ref{fig2}a), which shows the dominant configuration of the excitonic state.
We observe that the spins arrange antiferromagnetically along the $x$ direction, whereas they have a lower antiferromagnetic correlation along the $y$ direction.
This may lead to a spin arrangement with momentum $\mathbf{q}=(\pi,0)$ in the excitonic state and gives rise to the low-energy magnetic excitations as in the lowest-energy two-spinon excitation at $q=\pi$ of a Heisenberg chain.
We emphasize that anisotropic spin correlation is characteristic of the excitonic state and is the origin of the photoinduced low-energy magnetic excitations found in this paper.
This reasoning on anisotropic spin correlation is consistent with an independent numerical simulation of the spin and charge correlation in the excitonic state~\cite{Tokimoto2022},  reporting an anisotropic correlation between the $x$ and $y$ directions.

An increase of approximately 0.15~eV (1000~cm$^{-1}$ to 1500~cm$^{-1}$) in the low-energy spectral weight has been observed in recent trTMR experiments for YBa$_2$Cu$_3$O$_{6.1}$~\cite{Yang2020} when the material is driven by 1.5~eV pump pulse.
Thus, our finding that low-energy magnetic excitations below the two-magnon peak emerge when the Mott-gap edge region is driven by pump pulse is consistent with the experimental observations.
In the trTMR experiment, the polarization of pump photon was the (1,1) direction, which is different from the (1,0) direction in our case.
We confirmed that even in the (1,1) direction, low-energy excitations emerge in our calculation (see Supplementary Note ).
In real cuprates, second-neighbor hopping $t'_\mathrm{hop}$ cannot be neglected.
We also confirmed that the presence of $t'_\mathrm{hop}(=-0.25)$ does not change the emergence of low-energy magnetic excitations in trTMR (see also Supplementary Note).
To confirm our observations, we propose trTMR experiments that vary the pumping frequency where the low-energy weight is expected to be the maximum for the 1.5~eV pump pulse.

In summary, we have theoretically propounded photoinduced low-energy magnetic excitations below single- and two-magnon energies in a driven Mott insulator on a square lattice.
The intensity of the low-energy magnetic excitations is maximized when the frequency of the pulse is tuned to the absorption edge and optically allowed states with the $E$ presentation of the $C_{4\mathrm{v}}$ point group are excited.
The optically allowed states are excitonic states with a pair of holon and doublon, where the magnetic energy gains by forming a directional spin correlation.
Because the low-energy magnetic excitations are associated with the directional spin correlation in the excitonic states, we propose a concept of exciton-assisted magnetic excitations.
The proposed photoinduced low-energy magnetic excitations agree with the previously reported increase of low-energy weight in the trTMR spectrum~\cite{Yang2020}.
This theoretical prediction will be confirmed if the pumping frequency is varied in the trTMR and trRIXS experiments for insulating cuprates and iridates.
For three-dimensional antiferromagnetic Mott insulators, the same low-energy magnetic excitations are expected if one can photoexcite the Mott-gap edge, but its intensity will be small, since the directional spin correlation in the excitonic states is supposed to be weak due to smaller quantum spin fluctuations as compared with two-dimensional Mott insulator.
Calculating exciton-assisted magnetic excitations in three-dimensional Mott insulators will be a theoretical challenge in the future.

\section{Methods}
\label{Sec4}
We use a square-lattice periodic Hubbard cluster with $L=4\times4$ to calculate the time-resolved dynamical spin susceptibility and time-resolved two-magnon Raman susceptibility using the Lanczos-type exact diagonalization method.
The time-dependent wave function is given by $|\psi (t+\Delta t)\rangle =  e^{-iH(t)\Delta t} |\psi(t) \rangle$ with time step $\Delta t$.
Using Taylor expansion~\cite{Shirakawa2020}, we obtain $|\psi (t+\Delta t)\rangle = \sum_{n=0}^\infty|\varphi_n(t)\rangle$ with $|\varphi_n(t)\rangle=(-iH(t)\Delta t)^n/n!|\psi(t) \rangle$.
Each term is iteratively obtained by $|\varphi_{n+1}(t)\rangle=-iH(t)\Delta t/(n+1)|\varphi_n(t)\rangle$ starting from $|\varphi_0\rangle=|\psi(t)\rangle$.
We use $\Delta t=0.01$ and the summation with respect to $n$ is truncated if the norm $\langle\varphi_n(t)|\varphi_n(t)\rangle$ is smaller than a critical value such as $10^{-14}$.
For time-dependent dynamical quantities, we use Eq.~(\ref{O2}) with $\eta=0.2$, where we perform continuum-fraction expansions based on the Lanczos method starting from two initial states $|\psi(t) \rangle$ and $O_{-\mathbf{q}}|\psi(t) \rangle$. 

We also use an $L=6\times 6$ lattice with open boundary conditions to calculate the time-resolved dynamical spin susceptibility using tDMRG.
Expending the time-evolution operator using the spherical Bessel function $j_l(x)$ of the first kind and the Legendre polynomial $P_l(x)$ up to $M$-th order as $e^{-iH(t)\Delta t}\simeq C(t)\sum_{l=0}^M (-1)^l(2l+1)j_l(\delta t/\omega_\mathrm{s}(t)) P_l(H_\mathrm{s}(t))$, where the scaled Hamiltonian $H_\mathrm{s}(t)=\omega_\mathrm{s}(t)[H(t)-\lambda_\mathrm{s}(t)]$ with scaling parameters $\omega_\mathrm{s}(t)$ and $\lambda_\mathrm{s}(t)$ and where $C(t)$ is a normalization factor, we use two-target states, $|\psi (t)\rangle$ and $|\psi (t+\Delta t)\rangle$, in the DMRG procedure for a given $t$ to construct a basis set that can express wave functions in the time-dependent Hilbert space~\cite{Shinjo2021}.
Using the two-target time-dependent DMRG procedure, we can calculate the time-dependent quantities in Eq.~(\ref{O1}) with high accuracy.
We maintain 6000 density-matrix eigenstates in our tDMRG and use $M\sim20$.
In calculating Eq.~(\ref{O1}), we use a time increment of 0.02 and truncate time $s$ at $t+55$, and perform the integration as a discrete Fourier transform with $\eta=0.1$.
Figure~\ref{fig1}c shows the time-resolved dynamical spin susceptibility at $\omega_\mathrm{p}=20$, where a non-resonant condition results in very small energy absorption into the system because of $\omega_\mathrm{p}\gg U$.
Under such conditions, tDMRG with the 6000 eigenstates provides reasonable convergence in the spin susceptibility.
However, if $\omega_\mathrm{p}$ satisfies the resonance condition with energy absorption, the convergence worsens.

In an $L_x\times L_y$ lattice with open boundary conditions, we define momentum $\mathbf{q}=(q_x,q_y)$ as $q_{x(y)}=n_{x(y)}\pi/(L_{x(y)}+1)$ ($n_{x(y)}=1,2,\cdots,L_{x(y)}$) and $S^z_\mathbf{q}=2[(L_x+1)(L_y+1)]^{-1/2}\sum_i \sin(q_x i_x) \sin(q_y i_y) S_i^z$. 



\nocite{*}


\section{Acknowledgments}
This work was supported by QST President's Strategic Grant (QST Advanced Study Laboratory); the Japan Society for the Promotion of Science, KAKENHI (Grant Nos. 19H01829, JP19H05825, 21H03455, and 22K03500) from Ministry of Education, Culture, Sports, Science, and Technology, Japan; and CREST (Grant No. JPMJCR1661), Japan Science and Technology Agency, Japan.
Part of the computational work was performed using the supercomputing facilities in QST and the computational resources of the supercomputer FUGAKU provided by the RIKEN Center for Computational Science through the HPCI System Research Project (Project ID: hp210041).



\clearpage

\def\tred{\textcolor{red}}
\def\tblue{\textcolor{blue}}
\def\tcyan{\textcolor{cyan}}
\def\tgreen{\textcolor{green}}



\title{Supplementary Information\\Exciton-assisted low-energy magnetic excitations in a photoexcited Mott insulator on a square lattice}

\author{Kenji Tsutsui}
\affiliation{Synchrotron Radiation Research Center, National Institutes for Quantum Science and Technology, Hyogo 679-5148, Japan}

\author{Kazuya Shinjo}
\affiliation{Computational Quantum Matter Research Team, RIKEN Center for Emergent Matter Science (CEMS), Wako, Saitama 351-0198, Japan}

\author{Shigetoshi Sota}
\affiliation{Computational Materials Science Research Team, RIKEN Center for Computational Science (R-CCS), Kobe, Hyogo 650-0047, Japan}

\author{Takami Tohyama}
\affiliation{Department of Applied Physics, Tokyo University of Science, Tokyo 125-8585, Japan}

\date{\today}
             



\section*{Supplementary Note}
In the main text, we showed time-resolved two-magnon Raman (trTMR) spectrum for a half-filled Hubbard model with nearest-neighbor hopping $t_\mathrm{hop}$ and the on-site Coulomb interaction $U$ given by
\begin{equation}
H_0=-t_\mathrm{hop}\sum_{i\delta\sigma} c^\dagger_{i\sigma} c_{i+\delta\sigma}
 + U\sum_i n_{i\uparrow}n_{i\downarrow}, \tag{S.1}
\label{singleH}
\end{equation}
where $c^\dagger_{i\sigma}$ is the creation operator of an electron with spin $\sigma$ at site $i$, number operator $n_{i\sigma}=c^\dagger_{i\sigma}c_{i\sigma}$, $i+\delta$ represents the four nearest-neighbor sites around site $i$.
In insulating cuprates such as La$_2$CuO$_4$, second-neighbor hopping $t^\prime_\mathrm{hop}$ is necessary to describe its electronic states.
Therefore, it is important to investigate the effect of $t^\prime_\mathrm{hop}$ on trTMR.
The corresponding term is performed by
$-t_\mathrm{hop}^\prime\sum_{i\delta'\sigma} c^\dagger_{i\sigma} c_{i+\delta'\sigma}$,
where $i+\delta'$ represents the four second-neighbor sites around site $i$.
The Peierls substitution of an external electric field given by a pump pulse is performed by $c^\dagger_{i,\sigma} c_{i+\delta(\delta')\sigma} \rightarrow e^{-i\mathbf{A}(t)\cdot\mathbf{R}_{\delta(\delta')}} c^\dagger_{i,\sigma} c_{i+\delta(\delta')\sigma}$.
Here, $\mathbf{R}_{\delta(\delta')}$ is the vector from $i$ to $i+\delta(\delta')$, and $\mathbf{A}(t)$ at time $t$ is the vector potential given by
\begin{equation}
\mathbf{A}(t)=\mathbf{A}_0 e^{-(t-t_0)^2/(2t_d^2)} \cos[\omega_\mathrm{p}(t-t_0)],  \tag{S.2}
\label{A}
\end{equation}
where a Gaussian-like envelope centered at $t=t_0$ has temporal width $t_d$ and central frequency $\omega_\mathrm{p}$.

\begin{figure}[t]
\includegraphics[width=0.4\textwidth]{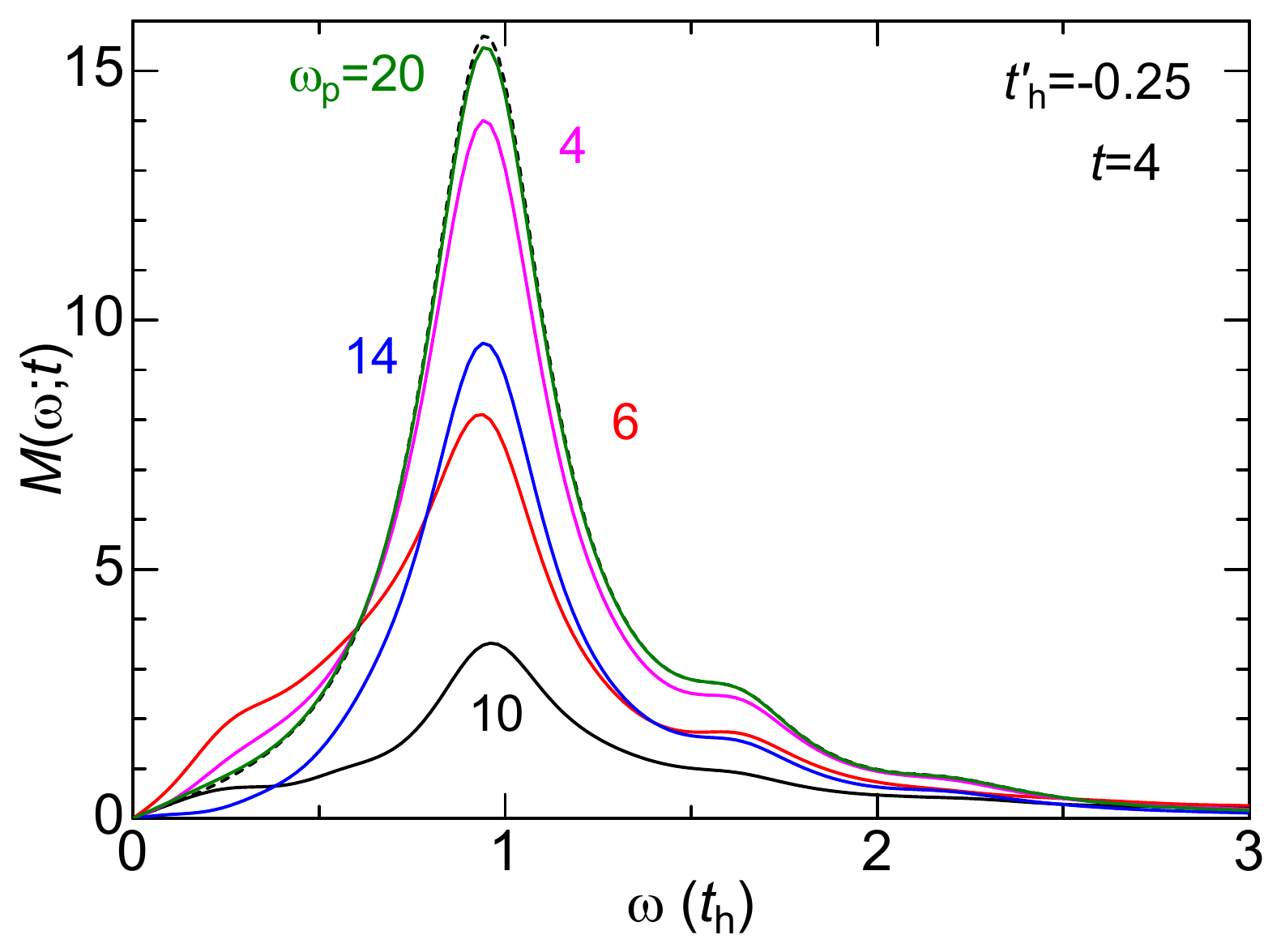}
\caption{\textbf{Pumping frequency dependence of time-resolved two-magnon Raman susceptibility after pumping.}
Time $t=4$ for a half-filled 4$\times$4 Hubbard lattice with on-site Coulomb interaction $U=10$ and second-neighbor hopping $t^\prime_\mathrm{hop}=-0.25$.
The dashed back line represents the results before pumping, and the purple, read, black, blue, and green solid lines represent the results at $\omega_\mathrm{p}$=4, 6, 10, 14, and 20, respectively.}
\label{fig1supp}
\end{figure}

As in the main text, we set $U=10$ by taking $t_\mathrm{hop}=1$ as the energy unit, and $t_0=0$ and $t_d=0.5$ by taking $1/t_\mathrm{hop}$ as the time unit.
In addition, we set $t^\prime_\mathrm{hop}=-0.25$~\cite{suppTsutsui2021}.
We then calculate B$_1$ trTMR susceptibility $M(\omega;t)$ after turning off a pump pulse having $\mathbf{A}_0=(A_0,0)$ with $A_0=0.5$ for a 4$\times$4 periodic lattice.
$M(\omega;t)$ was defined in the main text and numerically calculated using a Lanczos-type exact diagonalization technique.
Figure~\ref{fig1supp} shows the $\omega_\mathrm{p}$ dependence of $M(\omega;t)$ at $t=4$.
Similar to the case where $t^\prime_\mathrm{hop}=0$, as shown in Fig.~3 of the main text, the pump pulse suppresses the intensity of the two-magnon peak at $\omega=0.95$, which was observed in the trTMR experiment for photoirradiated YBa$_2$Cu$_3$O$_{6.1}$~\cite{suppYang2020}.
When $\omega_\mathrm{p}=6$, we find an enhancement in the low-energy spectral weight around $\omega=0.25$.
This is similar to the case without $t^\prime_\mathrm{hop}$ discussed in the main text.
These similarities clearly indicate that the presence of $t^\prime_\mathrm{hop}$ does not affect the emergence of low-energy magnetic excitations caused by photopumping.
We note that small shifts in both the two-magnon peak and photoinduced low-energy structure toward lower energy as compared with the case without $t^\prime_\mathrm{hop}$ (see Fig.~3 in the main text) can be attributed to the effect of magnetic frustration induced by second-neighbor superexchange interactions owing to finite $t^\prime_\mathrm{hop}$.

\begin{figure}[t]
\includegraphics[width=0.4\textwidth]{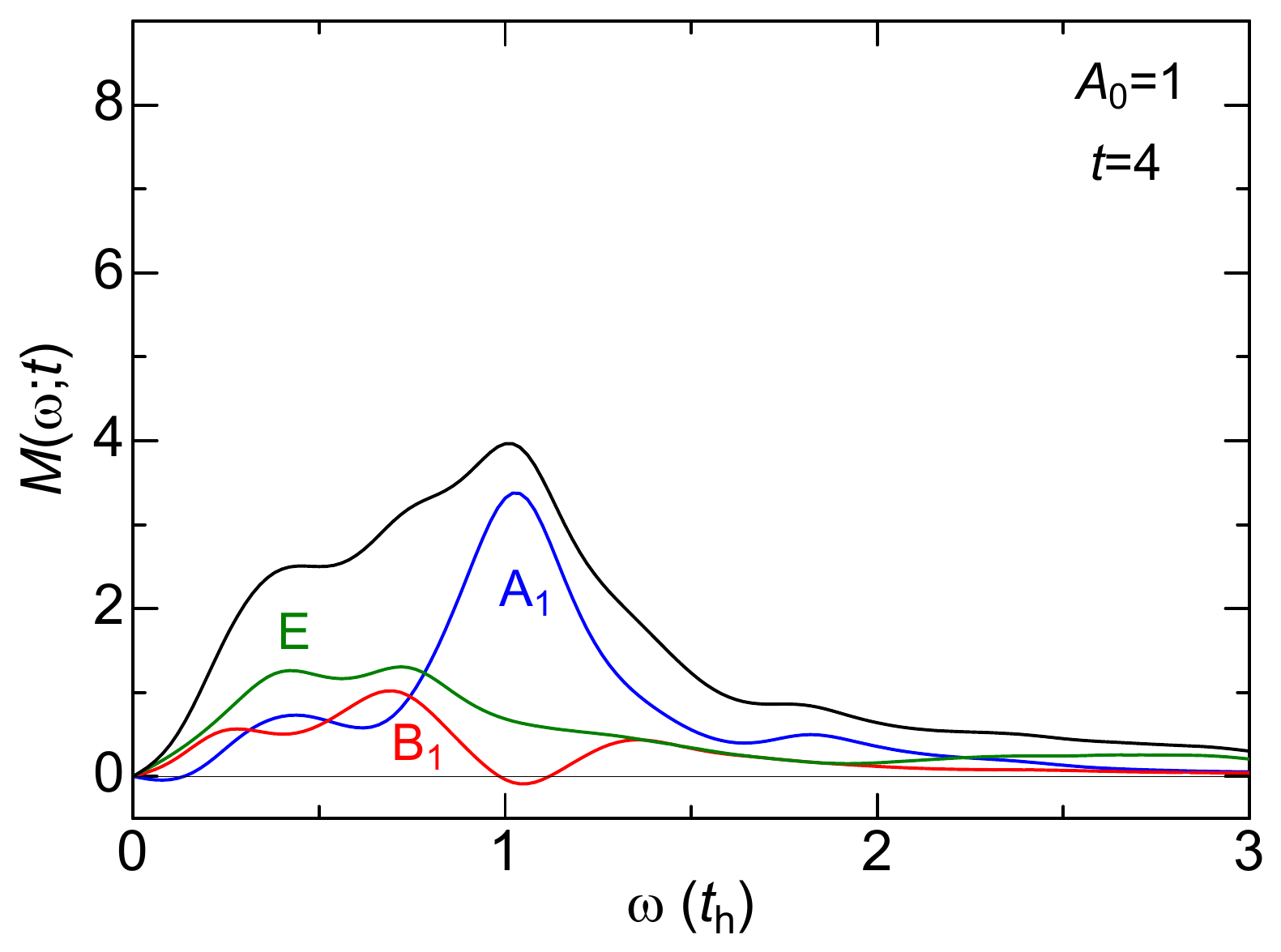}
\caption{\textbf{Symmetry decomposition of time-resolved two-magnon Raman susceptibility.}
Pumping frequency $\omega_\mathrm{p}=6$ and pumping amplitude $A_0=1$ at time $t=4$ for a half-filled 4$\times$4 Hubbard lattice with on-site Coulomb interaction $U=10$.
The black solid lines depict the sum of each contribution (the blue, green, and red solid lines depict the contributions of $A_1$, $E$, and $B_1$ of the $C_\mathrm{4v}$ point group, respectively.}
\label{fig2supp}
\end{figure}

In the main text, $A_0$ was set to $A_0=0.5$.
Increasing $A_0$ gives the change of spectral shape in the trTMR spectra~\cite{suppWang2018}.
To understand the effect of $A_0$, we show symmetry-decomposed $M(\omega;t)$ for $\omega_\mathrm{p}=6$ at $t=4$ when $A_0=1$ in Fig.~\ref{fig2supp}.
Note that a procedure in the symmetry decomposition was given in the main text.
Comparing the $A_0=1$ results with the $A_0=0.5$ results shown in Fig.~\ref{fig4} in the main text, we realize that the two-magnon peak is strongly suppressed and a single Lorentzian behavior near the maximum height disappears.
This change with increasing $A_0$ is related to the decrease of the $A_1$ component in the time-dependent wave function, which leads to the reduction of the two-magnon peak height.
We notice that the peak position in the $A_1$ component is the same as the two-magnon peak position before pumping.
If the pump pulse continues to the time region where the probe pulse comes in, the renormalization of antiferromagnetic exchange interactions is expected owing to the generation of the Floquet band~\cite{suppMentink2015}.
In the present calculation, however, the probe pulse comes in after the pump pulse finishes.
Therefore, there is no change of magnetic excitation energies due to the Floquet renormalization.
Instead, large contributions of the $E$ and $B_1$ components in the time-dependent wave function give rise to asymmetrical line shape of the trTMR susceptibility, as shown in Fig.~\ref{fig2supp}.
The asymmetric line shape leads to a low-energy shift in the centroid of spectral distribution.
Such a shift and the suppression of the two-magnon peak with increasing pump strength may explain experimentally observed pump-fluence dependence of the two-magnon peak~\cite{suppYang2020}.

\begin{figure}[tb]
\includegraphics[width=0.4\textwidth]{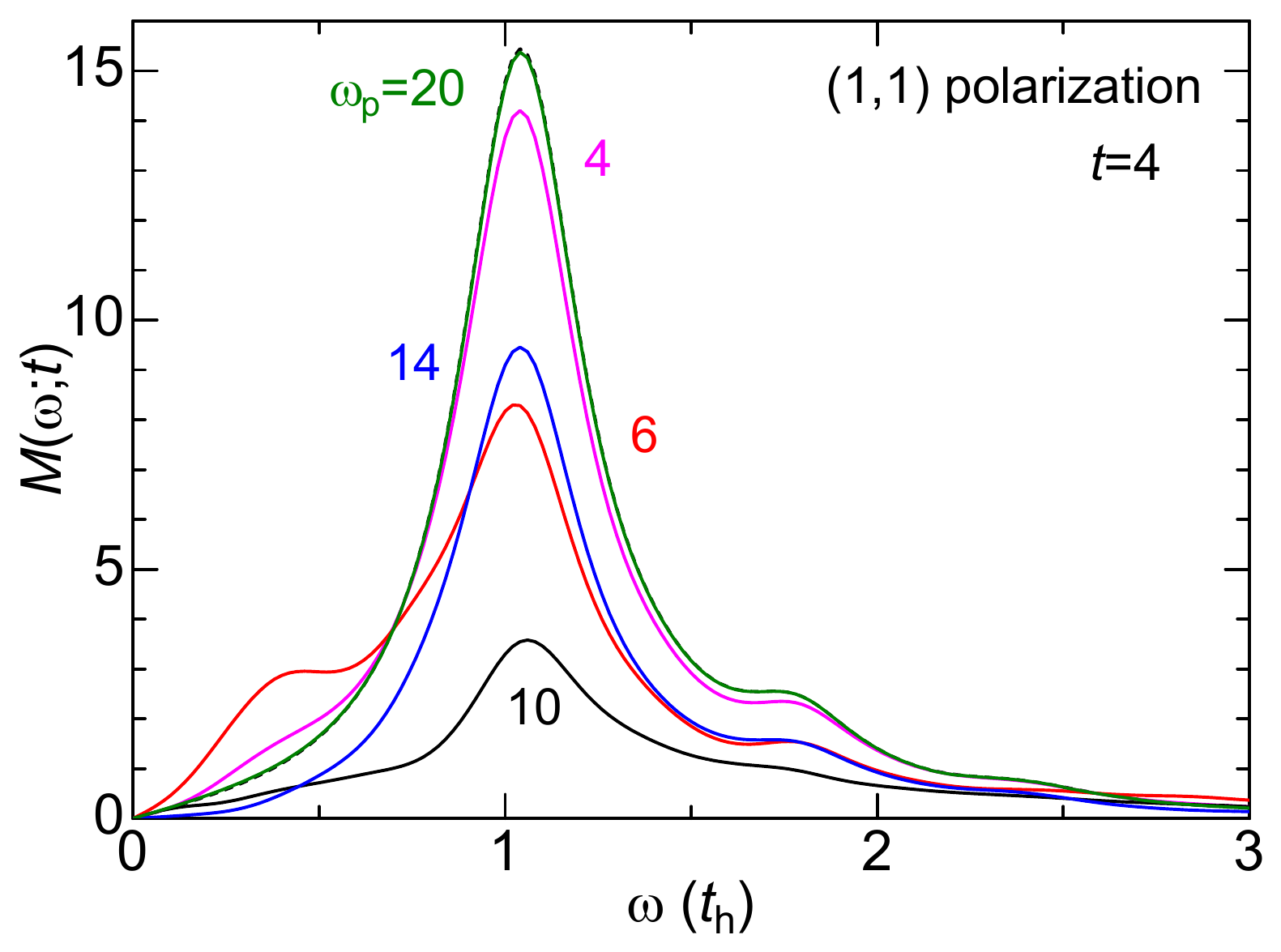}
\caption{\textbf{Pumping frequency dependence of time-resolved two-magnon Raman susceptibility after pumping.}
Time $t=4$ and the (1,1) polarization direction for a half-filled 4$\times$4 Hubbard lattice with on-site Coulomb interaction $U=10$ and second-neighbor hopping $t^\prime_\mathrm{hop}=0$.
The dashed back line represents the results before pumping, and the purple, read, black, blue, and green solid lines represent the results at $\omega_\mathrm{p}$=4, 6, 10, 14, and 20, respectively.}
\label{fig3supp}
\end{figure}

In the main text, the polarization vector $\mathbf{A}_0$ for the pump pulse was set along the (1,0) direction, that is, $\mathbf{A}_0=(A_0,0)$.
However, in the trTMR experiment for YBa$_2$Cu$_3$O$_{6.1}$~\cite{suppYang2020}, the polarization vector was selected the (1,1) direction.
To clarify whether the emergence of photoinduced low-energy magnetic excitations depends on polarization direction, we show the $\omega_\mathrm{p}$ dependence of B$_1$ trTMR susceptibility for $\mathbf{A}_0=(A_0/\sqrt{2},A_0/\sqrt{2})$ with $A_0=0.5$ in Fig.~\ref{fig3supp}. 
Photoinduced low-energy excitations exist at $\omega_\mathrm{p}=6$ as in the (1,0) direction.
Therefore, we conclude that the low-energy excitations are independent of polarization direction.
Thus, the observed one ranging from 1000cm$^{-1}$ to 1500cm$^{-1}$ in YBa$_2$Cu$_3$O$_{6.1}$ can be explained by the interpretation discussed in the main text.
   
\end{document}